\documentstyle[aps,multicol,epsf]{revtex}
\newcommand{\bqa}{\begin{eqnarray}}
\newcommand{\eqa}{\end{eqnarray}}
\newcommand{\bqs}{\begin{eqnarray*}}
\newcommand{\eqs}{\end{eqnarray*}}
\newcommand{\beq}{\begin{equation}}
\newcommand{\eeq}{\end{equation}}

\begin{document}
\title{Velocity correlations in granular materials}
\author{Tong Zhou}
\address{The James Franck Institute,
The University of Chicago, 5640 S. Ellis Avenue,  Chicago, IL 60637}
\date{\today}
\maketitle

\begin{abstract}
A system of inelastic hard disks in a thin pipe capped by hot walls is
studied with the aim of investigating velocity correlations between particles.
Two effects lead to such correlations: inelastic collisions help to build
localized correlations, while momentum conservation and diffusion produce
long ranged correlations. In the
quasi-elastic limit, the velocity correlation is weak, but it is still important
since it is of the same  order as the deviation from uniformity.  For
system with stronger inelasticity, the pipe contains a clump of particles
in highly correlated motion.   A theory with empirical parameters is
developed.  This theory is composed
of equations similar to the usual hydrodynamic laws of conservation of
particles, energy, and momentum.
Numerical results show that the theory describes the dynamics satisfactorily
in the quasi-elastic limit, however only qualitatively for stronger inelasticity.

\vspace{0.1in}
\end{abstract}
\begin{multicols}{2}

\section{Introduction}

A granular system normally consists of a large number of particles colliding
with one another and losing a little energy in each collision.   If such a
system is shaken to keep it in motion, its dynamics resembles that of fluids,
in that the grains move seemingly randomly. Considerable effort has been
devoted to the development of a continuum description for hydrodynamic
equations\cite{Goldshtein,GroR,BreD,Brey,Haf,JenS,JenR,Jen,HYH,SelG,GZB}.

Two approaches are employed by different authors.  One is to set up a
Boltzmann equation\cite{Goldshtein,GroR,BreD,Brey}, and then
to calculate hydrodynamic quantities by doing averaging with the
distribution function derived from the equation. In this case, the molecular
chaos assumption of the Boltzmann equation assumes zero correlations
between particles.  The other approach is to specify some hydrodynamic
quantities, and then write down the conservation equations for
them\cite{Haf,JenS,JenR,Jen,HYH,SelG,GZB}. Generally, there are three
equations: conservation of mass, balance of energy, conservation of
momentum.  The mass conservation is in the standard hydrodynamic form.
The momentum flux balance equation is in the form of Navier-Stokes equation for
fluid dynamics.  The energy ``conservation'' equation includes dissipation of
energy via collisions.

The failing of Liouville's theorem for granular systems\cite{zhou} casts doubts
on the applicability of conventional approaches to hydrodynamic equations.
In stead of writing down such equations based on unjustified assumptions,
studying the dynamics with as few assumptions as possible, and trying 
to develop a theory  closely connected to experiments, may be a less ambitious,
but more solid approach.

One major consequence of the usual hydrodynamic theories of fluids is the
Maxwell-Boltzmann distribution.  In the frame in which the average system
velocity is zero, this distribution implies no momentum correlations
whatsoever among different particles. This result is true for classical
particles independent of the strength of the inter-particle potential. In
contrast, however, granular materials commonly induce correlated collective
behaviors.  Think about the surface waves of vibrated
sand\cite{MelU}, or their convection patterns\cite{KJN}, (for a recent review, 
see\cite{JaeNB}.)  The
grains which take part in these collective behaviors do have correlated
velocities. We therefore ask: how important are these correlations and how
are they built up?

In this paper, we investigate the building up of correlations between
velocities of grains.  There are two mechanisms upon which correlations
can be built up.  One is the inelastic collisions between particles---after a
collision, the velocity difference between two particles is smaller than that
before the collision.  This is a local effect, and the correlation is short
ranged.  The other mechanism is from momentum conservation---the larger
the scale of a perturbation producing a localized average velocity, the slower
the perturbation decays\cite{GolZ}.  Fluctuations make the  system
non-uniform, so that there are localized clusters of particles all moving with
about the same velocity.  This correlation effect is a result of fluctuations
which are neglected in the usual hydrodynamic treatments.

There are hydrodynamic theories ignoring fluctuations which are consistent
with numerical results for weak inelasticities, but are quite inaccurate when
inelasticities are strong, see, e.g. \cite{LJH}.  These theories are attempts to
describe velocity fluctuations about some mean flow.  They work  fine in
quasi-elastic regime because correlations are small and negligible.  But
when  inelasticities are strong, collisions can bring groups of neighboring
particles to essentially the same velocity and thereby produce a correlated
motion which enhances the observable effect of any fluctuations in the
system.  We shall see this happen in our study.

The boundary conditions and system sizes independence of the essential 
characteristics of thermodynamic systems is one indication that thermodynamics
is a universal description.  However, this independence is lost in granular
systems.  We show that a universal description may still exist for the
unconserved modes of the dynamics.

\section{The thin pipe model}

\subsection{The system}

\begin{figure}
\vspace{-.8in}
\narrowtext
\centerline{\epsfxsize=9.6cm\epsfysize=9.6cm\epsfbox{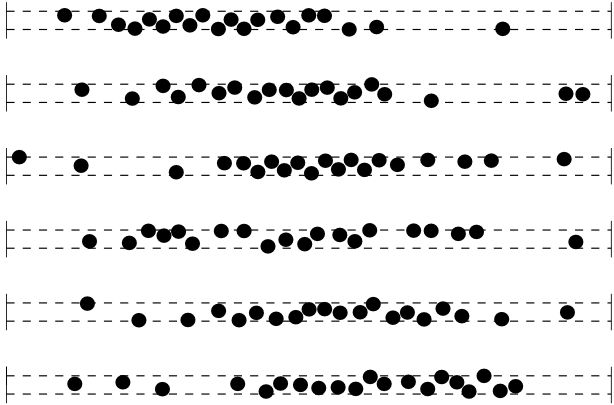}}
\caption{Snapshots of the thin pipe system.  The periodic side walls are
indicated by dashed lines.  The two end walls are energy sources kept at the
same temperature. Notice how most of the particles fall into a cluster, which
moves up and back through the system.}
\label{fg:pipe}
\end{figure}

To investigate the validity of a hydrodynamic description, we should study
the simplest situation which can show hydrodynamic behavior.  In the
elastic case, one dimensional systems have too many conservation laws and
do not show a fully ergodic or hydrodynamic behavior\cite{GroR,DLK}.
Here, we investigate a two-dimensional system in the form of a long thin
pipe (Fig.~\ref{fg:pipe}).  The grains confined in the pipe are all identical,
and the width of the pipe is set so that two grains cannot pass each other.
Thus the motion of grains is two dimensional to ensure ergodicity, while at
the same time we can order these  grains.   Pipe systems were studied 
before\cite{PosP}.  However, the no-passing condition enable
this thin pipe model to simplify
greatly both numerical and analytical calculations.  (A two-dimensional
version of this model is studied by Brey and Cubero\cite{Brey}.)

The two side walls are periodic---after leaving one side wall a particle comes
back through the other.  The distance between the side walls is chosen to be
$2.5$ times the radius of a particle.  This choice prevents any passing. Two
end walls are energy sources, and are kept at the same temperature.

For a
thermodynamic system, the bulk properties should not depend on the
details of boundary conditions.  However, for some granular properties,
boundary conditions can be quite important\cite{GroR,KJN}.  We employ
two different boundary conditions in the numerical calculations: In both
cases, when a particle hits an end wall the direction of its motion is turned
around, and the particle is returned to the system.      In the {\em fixed
speed} boundary condition, the  returned particles move away from the wall
with a unit speed.  Alternatively, in the {\em Boltzmann} boundary
condition the returning speed is picked from a distribution
$P(u)=2ue^{-u^2}$\cite{GroR}.  All figures describe simulations with 
the {\em Boltzmann} boundary conditions unless otherwise specified.

\subsection{The parameters and variables}

We use the simplest model:  non-rotating particles.   After a collision, the
radial relative velocity changes sign, and decreases by a factor of the
restitution coefficient $r$, with $0<r<1$. In the collision, the other
components of the velocities are  unchanged.  Thus, $r=1$ is for elastic
particles, and $r=0$ for extremely inelastic particles. We also define
$\epsilon\equiv 1-r$.

The coordinate in the problem is an index, $i$, which indicates the position
of the particle.
Suppose there are $N=2n$ particles in the thin pipe.  They are ordered as
$$-n, -(n-1),\cdots ,-1,0,1,\cdots ,(n-1).$$
By using the particle number, $i$, as our coordinate,  we
take advantage of the `no-passing' property of the thin pipe and thereby get
a Lagrangian description of the system.

Let us denote the velocity of the $i$th particle as $\vec{u}_i$, the relative
velocity between the $i$th and the $(i+1)$th particle as
$\vec{v}_i\equiv\vec{u}_{i+1}-\vec{u}_i$, and the velocity of the center
of mass as $\vec{u}$, the velocity of the $i$th particle with respect to the
center of mass as $\vec{u}_i^r\equiv\vec{u}_i-\vec{u}$.  Let us assume
the pipe is along the $x$ direction.  Then the $x$-component of the
velocities are special and we denote them as $u_i$ and $v_i$. We use an
over-line
notation to indicate the root mean square ({\em rms}) value of some quantities.

We propose a method to calculate profiles of various quantities
throughout the system and the velocity correlations. (A  {\em profile} is
a plot of the value of some averaged quantity as a function of the particle number
variable $i$.)  Instead of the strongly correlated velocities $\vec{u}$'s, we
study the relative velocities of neighboring particles, $\vec{v}$.
In using $\vec{v}$ we focus our attention on the
relative motion of the particles and away from their collective and correlated
motion.

There are four parameters which will describe our system, the particle number,
$N$, the pipe length $L$, width $W$, and the
inelasticity $\epsilon$.  Of course $\epsilon$ measures the total
amount of inelasticity in one collision. In a system with many particles, the
effect of the inelasticity is enhanced by the correlation effects. For this
reason, we expect two combinations of $N$ and $\epsilon$ to be important. The
product $N \epsilon$ measures the total amount of inelasticity in the system.
For a one dimensional system, imagine a particle with a large velocity
hitting a
group of $n$ particles, sitting almost at rest. The added momentum will cascade
down the group until at the end of the line the transmitted momentum will be
diminished by a factor $\exp(-n \epsilon)$.  In addition,  a previous
calculation\cite{GZB} showed that dissipation of energy led to a gradual decay
of temperature in the form of an exponential of $-c\sqrt{\epsilon}n $ where $c$
is a constant.  Thus we expect a dip in temperature determined by the
combination of parameters, $\sqrt{\epsilon}N$. Changing the remaining
parameters, $L$ and $W$, will
only modify some numerical factors in the theory---but will not change the
qualitative behavior of the system.

The system showed in Figure~\ref{fg:pipe} contains both low density and
intermediate density regimes.  There are some complication in such systems
because of different geometrical factors for different density
regimes\cite{GZB}.  To avoid such complication and focus on the
dynamics of the system, we carry out our numerical calculations only for
systems with extremely high density, where the typical spacing between
neighboring particles is about $2\%$ of the radius of a particle; or for
systems with extremely low density, where the spacing at the highest
density region of the system is about $10$ times the radius.  The essential
characteristics of the dynamics are independent of density regimes.

\subsection{The steady state}

This system can reach a statistical steady state.
In this state, the particles move fast near the hot
walls, and the density is low there.  Towards the center of the system, the
density is higher.
For quasi-elastic situations, the system is relatively uniform; but for
stronger inelasticities, the particles near the center
can form a cluster and move with about the
same velocity.  The cluster was seen and understood in
previous calculations\cite{GZB,GolZ}.  The relative motion of particles is
reduced by the inelastic collisions between them. In fact,  when $N\epsilon$ is
large, the relative motion can be very small and then momentum conservation
causes that each particle in the cluster
has about the same velocity which is just the mean
velocity of the cluster.

\begin{figure}
\narrowtext
\epsfxsize=\hsize
\vspace{-.2in}
\epsfbox{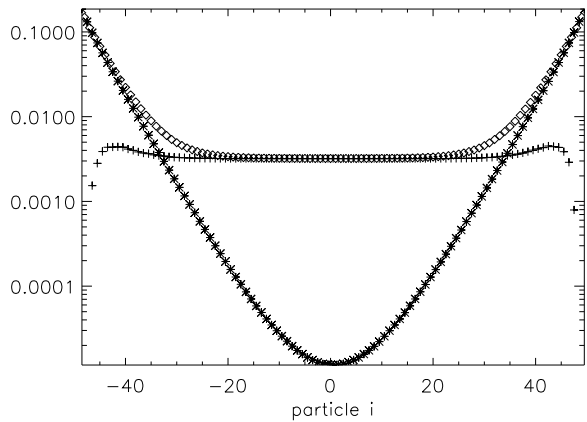}
\caption{Profiles of a low density system of 100 particles with $r=0.94$,
just above the critical value for inelastic collapse $r_c$.  $\diamond$ is for
$\langle u_i^2\rangle$, $\ast$ for $\langle v_i^2/2\rangle$, and $+$ for
$\langle u_i u_{i+1}\rangle$. }
\label{fg:profiles}
\end{figure}

Figure~\ref{fg:profiles} shows a plot of some profiles in an inelastic
situation with two hot walls.  Notice that the profile of $\langle
u_i^2\rangle$
 has a flat region at the center.  This was seen before \cite{Brey}.  That
flattening occurs because the central particles almost always fall within a
cluster, and the cluster moves with a large average velocity but small relative
velocities.
The plot of  $\langle v_i^2\rangle$ indeed shows that the relative velocity
decreases to a very small value near the center of the system.   This decay in
 $\langle v_i^2\rangle$ is roughly what we might expect from a simple
hydrodynamic description, in which one balances energy flux with
dissipation\cite{GZB}.  The hydrodynamics then gives an $i$-dependence which is
a superposition of growing and decaying exponentials. That theory is in some
sense a mean field theory which ignores the correlations between
velocities.
In much of
what we do, delicate and long-range correlations effects will be very
important for the behavior of $\vec{u}$'s but less important for the
$\vec{v}$'s.
In fact we shall see that the rms of $v_i$ obeys
\beq
\frac{\partial^2}{{\partial i}^2} \bar{v}_i =b^2 \bar{v}_i,
\label{eq:v}
\eeq
where $b^2$ is proportional to $\epsilon$ for small
values of the inelasticity.  The solution to the equation is
\beq
\bar{v}_i =  \bar{v}_0 \cosh(bi).
\label{eq:sol}
\eeq
Equations (\ref{eq:v}) and (\ref{eq:sol}) describe a situation in
which heat conduction balances against energy dissipation.

On the other hand, the large degree of correlation between $u_i$ and
$u_{i+1}$ is quite unexpected.  No such correlation occurs in the usual
statistical mechanics. This kind of correlation effect is not directly
contained in any hydrodynamic equations.  As we shall see, it is a result of
fluctuations not usually included in hydrodynamics.

\begin{figure}
\narrowtext
\epsfxsize=\hsize
\vspace{-.2in}
\epsfbox{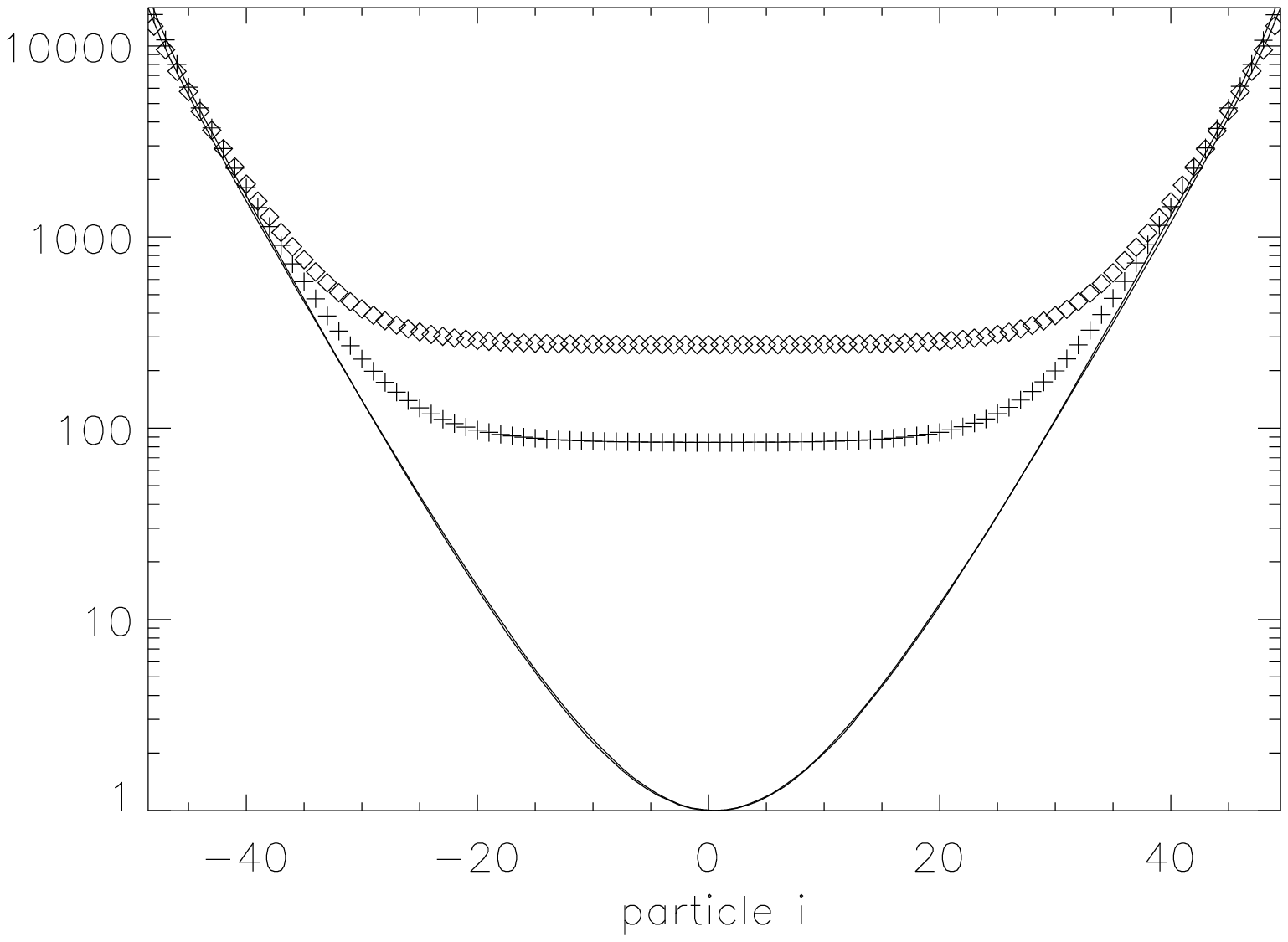}
\caption{Profiles of $\langle v^2_i/2\rangle$ and $\langle u^2_i/2\rangle$ for
two different boundary conditions.
The system has low density $100$ particles and $r=0.94$.
Each profile is rescaled by changing the scale of velocity so that $\langle
v^2_0/2\rangle =1$.  There are two lines, which nearly overlap each other,
describing $\langle v^2_i/2\rangle$.  The $\diamond$ is for $\langle
u^2_i\rangle$ with the {\em
Boltzmann} boundary condition, and the $+$ is for $\langle u^2_i\rangle$
with {\em fixed speed} boundary condition.  These two profiles are
very different.  }
\label{fg:boundary}
\end{figure}

Boundary conditions are often important for granular systems.
Figure~\ref{fg:boundary} shows the effects of boundary conditions.
We see $\langle u_i^2\rangle$ depends sensitively on boundary conditions.
In contrast, after a rescaling,  $\langle v_i^2\rangle$ is nearly
independent of
boundary conditions. There is no similar rescaling which can make the profiles
for  $\langle u^2_i\rangle$ overlap.

\subsection{Correlated motion and random motion}

Since
\beq
\langle v_i^2\rangle=\langle u_i^2+u_{i+1}^2\rangle-2\langle u_i
u_{i+1}\rangle,
\label{eq:div1}
\eeq
when the correlation $\langle u_i u_{i+1}\rangle$ is weak, we
simply have $\overline{v}_i^2=2\overline{u}_i^2$, assuming a weak
$i$-dependence.  But when correlation is
strong, the relation between $\overline{v}_i^2$ and
$\overline{u}_i^2$ is quite different. We shall study that difference in
detail.  From the mechanism described above, we know near the center,
$\overline{u}_i^2$ is roughly constant, independent of $i$, as a consequence of
the motion of the cluster. Conversely,  $\overline{v}_i^2$ will vary because of
energy dissipation.  In our considerations, we shall focus upon
${v}_i^2$, which has an average which can be interpreted as a local
temperature.  We argue that ${v}_i^2$ is a more relevant variable than
$u_i^2$, since to a large extent, it determines the collision rate, and the
effect
of a collision. In addition, $\overline{v}_i$ behaves as predicted by the
simple
hydrodynamics theory, it decays exponentially, and forms a hyperbolic cosine
curve as a function of $i$. Conversely, $\overline{u}_i$ is produced by
subtle correlation effects.

We can also write (\ref{eq:div1}) in the form
\beq
\frac{1}{2}\langle u_i^2+u_{i+1}^2\rangle=\frac{1}{2}\langle v_i^2\rangle+
\langle u_i u_{i+1}\rangle.
\label{eq:div2}
\eeq
The term on the left hand side describes the total motion with respect to
the lab frame, the
second term on the right hand side describes the correlated motion between
particle $i$
and particle $i+1$, and the first term on the right
hand side describe the
random relative
motion between neighboring particles.  So put this into words,
$$\left(\begin{array}{c}
\mbox{total}\\ \mbox{motion}\end{array}\right)
\quad = \quad
\left(\begin{array}{c}\mbox{random}\\ \mbox{motion}\end{array}\right)
\quad + \quad
\left(\begin{array}{c}\mbox{correlated}\\ \mbox{motion}
\end{array}\right).$$
The first term on the right can be interpreted as a temperature; the second as
a result of the correlated motion of the two particles. In this way, we see
that the ratio
\beq R_i =\frac{\langle v_i^2\rangle}{\langle
u_i^2+u_{i+1}^2\rangle}=
\frac{\mbox{random motion}}{\mbox{total motion}},
\label{eq:rat}
\eeq
indicates the amount of correlation in the motion.  When
the inelasticity is weak, the velocity correlations are also weak, and this
ratio is very close to unity.
For strong inelasticity, where correlations are strong, this ratio can be
very small.

\subsection{PDF's of velocities}

\begin{figure}
\epsfxsize=\hsize
\epsfbox{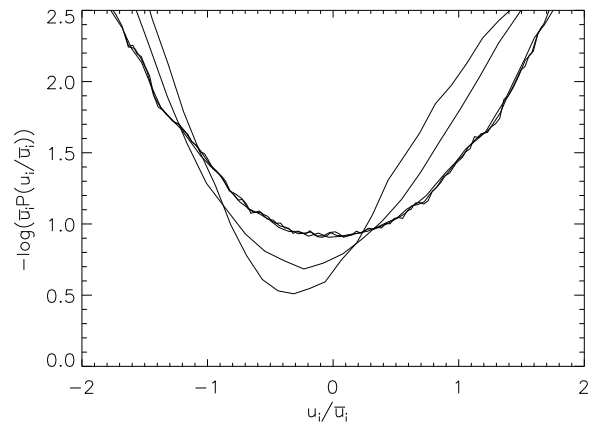}
\caption{Time-averages probability for $u_i$ in a low density system with
$N=100$, $r=0.94$.  Five such curves are shown, which are for
$i=0,-10,-20, -35,-45$.  The first three in this list are close to Gaussian
and they lie almost on top of one another. The other two, are quite  different.}
\label{fg:udis}
\end{figure}

\begin{figure}
\epsfxsize=\hsize
\epsfbox{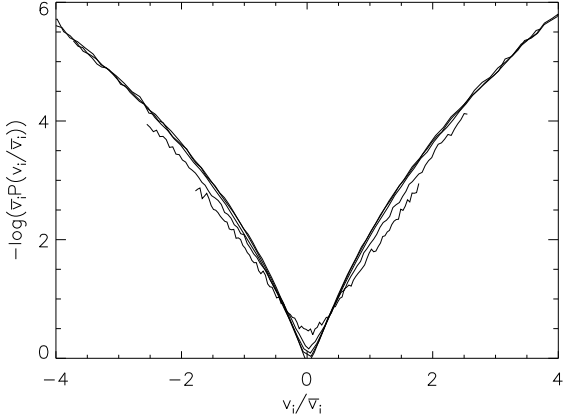}
\caption{Probability distribution for relative velocities. The calculation
is done as a time
average for a low density system with$N=100$,
$r=0.94$ and $i=0,-10,-20, -35,-45$.  the PDF's for $v_i$'s collapse into a
single curve after
a rescaling. Once again the three curves for particles inside the cluster
fall on top of one
another, while the others are slightly different. }
\label{fg:vdis}
\end{figure}

The probability distribution functions (PDF) for $u_i$ and  $v_i$
provide considerable additional insight into the nature of the system. See
figures  (Fig.~\ref{fg:udis}) and (Fig.~\ref{fg:vdis}). In these figures,
the variables are normalized to give each PDF the same variance.

In the PDF plots for $u_i$, we see a fundamentally Gaussian behavior inside
the cluster.  Outside the cluster, the part of the curve shown is Gaussian
but there is a strong high velocity tail.

In contrast, the PDF plots for  $v_i$ show a structure which is
essentially the same inside and outside the cluster. 
Thus, all over the system, the $v's$
behave in the same way, but this behavior is quite non-trivial.

We will use the constancy of the PDF of $v_i$ (in the whole system) and $u_i$
(in the interior of the system) to develop our theoretical model.

\section{Motion in the center of mass frame}

Figure~\ref{fg:boundary} suggests that we can decompose the dynamics
of the system into two parts:  I) the motion of grains in the center of mass
frame and II) the motion of the center of mass itself. Part I is
independent of boundary conditions and all the effects of
boundary conditions are attributed to Part II\@.  Part I is described in
terms of the variables $\vec{v}_i$ which may be considered to be weakly
correlated with one another.   Part II involves variables $\vec{u}_i$, and
strong
correlations among the variables.  In this section, we focus our attention upon
the effects of conservation laws upon the system, and particularly on motion
of Part I.

\subsection{Theoretical calculation}

Since the number of degrees of freedom of Part I
is equal to the number of $\vec{v}_i$'s,
this part of motion can be described
in terms of $\vec{v}_i$'s.  So the problem can be solved in two steps:
the $rms$ of $\vec{v}_i$'s and the correlations between $\vec{v}_i$'s.
Our interests in the variable $\vec{v}_i$'s are also based on the numerical
results showed in the previous section that the profile of $\overline{v}_i$
is, in accordance to hydrodynamics theory and Equation~(\ref{eq:sol}),
a hyperbolic cosine function of
$i$, plus its weak dependence on the boundary conditions---these
suggest that $\vec{v}_i$ can form the basis of a solution to some simple
hydrodynamics equations.

\subsubsection{Profile of $\overline{v}_i$}

{\bf Collisions}

For the steady state, mass conservation is reduced to a trivial statement that
$\langle\vec{u}_i\rangle=0$ and $\langle\vec{v}_i\rangle=0$.  The
momentum and energy transfer between particles are results of collisions
between them.  So to investigate momentum and energy conservation, we
study the effects of a single collision first.

Let us consider a collision between the $i$th and the $(i+1)$th particle,
during which $\vec{u}_i$, $\vec{u}_{i+1}$, and $\vec{v}_i$ change to
$\vec{u}_i'$, $\vec{u}_{i+1}'$, and $\vec{v}_i'$ respectively.
According to the inelastic collision rule,
\bqa
\vec{u}_i' & = & \vec{u}_i+\frac{1+r}{2} \hat{n} {v}_{i,n},
\label{eq:coll1}\\
\vec{u}_{i+1}'&=&\vec{u}_{i+1}-\frac{1+r}{2} \hat{n} {v}_{i,n},\\
\vec{v}_i'&=&\vec{v}_i-(1+r)\hat{n}{v}_{i,n},
\label{eq:coll3}
\eqa
where $\hat{n}$ denotes a unit vector, pointing in the direction of the line
of centers at the point of collision while ${v}_{i,n}$ is the component of
$\vec{v}_{i}$ in that direction.

{\bf Pressure}

The collision described above
results in a change in the momentum of particle $i+1$,
$$\vec{u}_{i+1}'-\vec{u}_{i+1}=-\frac{1+r}{2}\hat{n}{v}_{i,n}.$$
In a long time interval $t$, the momentum change of particle $i+1$ from
collisions between particle $i$ and particle $i+1$ is
\beq
P_i W t= -\frac{1+r}{2}\sum^{(i)}(\hat{n}\cdot\hat{x}) v_{i,n},
\label{mom}
\eeq
where $\sum^{(i)}(\cdots )$ is summation over all the collisions between
the $i$th particle and the $(i+1)$th particle. The $x$ component in
Equation~(\ref{mom}) is the direction along the pipe.

In writing Equation~(\ref{mom}) we have identified the rate of momentum
transfer
from particle $i$ to particle $i+1$ as an average pressure, $P_i$ times the
pipe-width, $W$, while $t$ is the time for the summation.  We shall be
dealing a
lot with sums over collisions as in equation (\ref{mom}). To understand them,
we
should realize that  $\sum^{(i)}(\cdots )/t$ can be written as the rate of
collisions between $i$ and $i+1$, $c_i$,  times an average over collisions
$\langle\cdots\rangle_i$ of this type.  Notice that the average over collisions
is very different from the time-average $\langle\cdots\rangle$.
For example, $\langle v_i\rangle$ must be zero
in any steady state situation. However, since $v_i$ must be negative for a
collision to
occur, then $\langle v_i\rangle_i$ must be negative.

Now go back to Equation~(\ref{mom}).  For the steady state, the momentum flux
must
be a constant, so this summation over a long time interval must be
independent of
$i$. Thus the momentum conservation law has the consequence that the pressure
as defined by Equation~(\ref{mom}) is independent of $i$. So the equation for
momentum conservation in our system takes the form
\beq
-\frac{1+r}{2 W t} \sum^{(i)} (\hat{n}\cdot\hat{x}) v_{i,n}
= -\frac{1+r}{2 W }
c_i \langle (\hat{n}\cdot\hat{x})v_{i,n}\rangle_i = P.
\label{eq:mb}
\eeq

The distribution functions for relative velocity
only depend weakly on $i$ (Fig.~\ref{fg:vdis}).  Thus,
all components and averages of $\vec{v}_i$ vary in proportion to one
another as $i$ is varied. As $i$ approaches the center of the system, the
typical value of the momentum transfer per collision declines in proportion to
$\overline{v}_i$. Then, by Equation~(\ref{eq:mb}) the collision rate
increases by
going inversely as the relative velocity.

Using the same arguments,  we can also understand the
pressure-definition, Equation (\ref{eq:mb}), in a familiar form. Pressure
is a flow of momentum
per unity area per unit time. One kind of flow involves transfer of
momentum from the $i$th     
particle to the next one. The collision rate is of the order of
$\overline{v}_i/l_i$, where
$l_i$ is the mean spacing between the two particles. During each collision,
the average
momentum transfer is proportional to $\overline{v}_i$.  From these two
facts, the momentum flux
is proportional to $\overline{v}_i^2/l_i$.  The average of relative velocity
squared is the temperature, $T$, while $1/(W l_i)$ is the density, $\rho$.
This result is then
in the familiar form,  $P=\rho T$.  This identification is an order of
magnitude argument.  For calculations, we use the exact result,
Equation~(\ref{eq:mb}).

{\bf Energy Balance}

Now let us study the effects of this collision on the energy balance.  The
energy transfer to particle $i$ and particle $i+1$ are, respectively,
\bqs
\vec{u}_i'^2-\vec{u}_i^2&=&\frac{1+r}{2}\left({u}_{i+1,n}^2-{u}_{i,n}^2
\right)-\frac{1-r^2}{4}{v}_{i,n}^2,\\
\vec{u}_{i+1}'^2-\vec{u}_{i+1}^2&=&\frac{1+r}{2}\left({u}_{i,n}^2-{u}_{i+1,n}^2
\right)-\frac{1-r^2}{4}{v}_{i,n}^2,
\eqs
and the energy dissipation is
$$\left[\left(\vec{u}_i^2+\vec{u}_{i+1}^2\right)
-\left(\vec{u}_i'^2+\vec{u}_{i+1}'^2\right)\right]
=\frac{1-r^2}{2}{v}_{i,n}^2.$$

We can form an energy conservation equation by balancing the energy
dissipation with the difference of the energy flux.  However, the above
expressions involve $\vec{u}_i$'s which are correlated and do not
belong to the motion in the center of mass frame.  To find a consistent
description, we want to express this conservation in terms of $\vec{v}_i$'s.
Because the essential dynamic process is determined by the collision rule, 
Equations (\ref{eq:coll1}-\ref{eq:coll3}), an equation describing the balance
of a quadratic form of $\vec{v}_i$'s will incorporate the energy conservation.

In fact, we have, from (\ref{eq:coll1}-\ref{eq:coll3}),
\bqs
\vec{v}_i'^2-\vec{v}_i^2 & = & -(1-r^2){v}_{i,n}^2,\\
\vec{v}_{i+1}'^2-\vec{v}_{i+1}^2 & = &
(1+r){v}_{i,n}{v}_{i+1,n}+
\frac{(1+r)^2}{4}{v}_{i,n}^2,\\
\vec{v}_{i-1}'^2-\vec{v}_{i-1}^2 & = &
(1+r){v}_{i,n}{v}_{i-1,n}+
\frac{(1+r)^2}{4}{v}_{i,n}^2.
\eqs
For the steady state, the total change in $v_i^2$ should vanish,
$$\sum^{(i)}(\vec{v}_i'^2-\vec{v}_i^2)+
 \sum^{(i+1)}(\vec{v}_i'^2-\vec{v}_i^2)+
\sum^{(i-1)}(\vec{v}_i'^2-\vec{v}_i^2)=0,$$
or equivalently,
\bqa
&-&(1-r)\sum^{(i)}{v}_{i,n}^2 \nonumber\\
&+&\sum^{(i+1)}({v}_{i,n}{v}_{i+1,n}
+\frac{1+r}{4}{v}_{i+1,n}^2) \nonumber\\
&+&\sum^{(i-1)}({v}_{i,n}{v}_{i-1,n}
+\frac{1+r}{4}{v}_{i-1,n}^2)=0.
\label{eq:eb}
\eqa
The first term is from energy dissipation, while the other terms take
the form of energy transfer.

{\bf Profile of $\overline{v}_i$}

We wish to simplify our energy-flow equation by reducing it to an
equation for $\langle v_{i,n}^2\rangle_i$.  However, correlations between
$\vec{v}_i$ and
$\vec{v}_{i+1}$ appears in (\ref{eq:eb}). We must eliminate these terms.
For an
elastic uniform system, this correlation takes a simple form,
\bqa
\langle v_iv_{i+1}\rangle&=&
\langle (u_{i+1}-u_i)(u_{i+2}-u_{i+1})\rangle\nonumber\\
&=&-\langle u_{i+1}^2\rangle\label{eq:ec}\\
&=&-\frac{1}{2}\overline{v}_i\overline{v}_{i+1}\nonumber
\eqa
In the elastic case, it is equally true for the usual
time-weighted average or for the collision weighted average, as
\bqa
&&\sum^{(i+1)}({v}_{i,n}{v}_{i+1,n})+\sum^{(i-1)}(
{v}_{i,n}{v}_{i-1,n}) \nonumber\\
&=&-\frac{1}{2}
(n_{i+1}^cv_i^cv_{i+1}^c
+n_{i-1}^cv_i^cv_{i-1}^c),
\label{vlaw}
\eqa
where $v_i^c\equiv \sqrt{\sum^{(i)}\vec{v}_{i,n}^2/n_i^c}$,
and $n_i^c$ is the total number of collisions between particle $i$ and
particle $i+1$, $n_i^c=c_it$.
As defined here, $v_i^c$ is an collision average of $\vec{v}_i$ just before
collisions.

Equation (\ref{vlaw}) has scalars on the left and right hand side.
There are
corrections to this relation for inelastic particles and when there is
a spatial variation in the averages.  The corrections must be scalars and of
order ${\overline{v}_i}^2$. One correction is of order of the order of
$\epsilon
{\overline{v}_i}^2$. In the other correction,  $\frac{d^2}{di^2}$ is
applied to
${\overline{v}_i}^2$. However, in virtue of the result in Equation (\ref{eq:v})
these two terms are really the same.  Consequently, we need only one of these
two corrections.  We write
the resulting structure in an even parity form as
\bqa
&&\sum^{(i+1)}({v}_{i,n}{v}_{i+1,n})+\sum^{(i-1)}(
{v}_{i,n}{v}_{i-1,n})\nonumber\\
&=&-\frac{1-a_1\epsilon}{2}
(n_{i+1}^cv_i^cv_{i+1}^c
+n_{i-1}^cv_i^cv_{i-1}^c),
\label{eq:a1}
\eqa

Now Equation (\ref{eq:a1}) can be used to transform equation
 (\ref{eq:eb})  into the form
\bqs
-\epsilon n_i^c{v_i^c}^2+\frac{1+r}{4}(n_{i+1}^c{v_{i+1}^c}^2
+n_{i-1}^c {v_{i-1}^c}^2)\\
-\frac{1-a_1\epsilon}{2}(n_{i+1}^c v_i^cv_{i+1}^c
+n_{i-1}^cv_i^c v_{i-1}^c)=0,
\eqs
or, since $n_i^cv_i^c$ is a constant independent of $i$, (see Equation
(\ref{eq:mb})), we find a heat flow equation
$$(6-4a_1)\frac{\epsilon}{1+r}\overline{v}_i=
\overline{v}_{i+1}-2\overline{v}_i+\overline{v}_{i-1}.$$
In writing the last structure we have noticed that the different kinds of
collision averages all have the same $i$-dependence.
Now we can phrase our
result in a continuum form
$$(6-4a_1)\frac{\epsilon}{1+r}\overline{v}_i=
\frac{d^2\overline{v}_i}{di^2}.$$
In this way, we obtain
\beq
\overline{v}_i=\overline{v}_0\cosh (bi).
\label{eq:pro2}
\eeq
where
\beq
b^2=(3-2a_1)\epsilon.
\label{eq:b}
\eeq

\subsubsection{Correlations between velocities}

Correlations between $v_i$'s are short ranged.  Let us only consider the
nearest
neighbor correlation.  When there is no dissipation, the only non-vanishing
correlation of the $v_i$'s is the nearest neighbor average of Equation
(\ref{eq:ec}).  For $r<1$, there is a small correction to that relation.
 Just as before, (see Equation (\ref{eq:a1})), we
write an equation for the average of a nearest neighbor product in the same
form as in
the elastic case, but with a correction proportional to $\epsilon$:
\beq
\langle v_i v_{i+1}\rangle=-\frac{1-a_2\epsilon}{2}\overline{v}_i
\overline{v}_{i+1},
\label{eq:a2}
\eeq
where the averages are time average.

This assumption, with the profile of $\overline{v}_i$ determined above,
completes a
description of the motion of grains in the center of mass of frame, i.e.\
Part I of the
dynamics described before.  As an example, let us calculate the
correlations between $u_i^r$'s, the velocities of particles in the center of
mass frame.  To illustrate the similarity between this part of the dynamics
and conventional thermodynamics, i.e., the independence of boundary conditions 
and system sizes, we consider the center of mass frame of the $2m$ particles
at the center of the system.  
Keep in mind that rather than fixed, $m$ can be treated as a variable
in the following calculation.

Express $u_i^r$ in terms of $v_i$'s,
$$
u_i^r=-\frac{1}{2m}\left[\sum_{j=i}^{m-1}(m-j)v_j-\sum_{j=-(m-1)}^{i-1}(m+j)v_j\right].
$$
So
$$
\frac{2u_i^ru_{i+1}^r}{{u_i^r}^2+{u_{i+1}^r}^2}=\frac{A-m^2v_i^2}{A+m^2v_i^2},
$$
where
$$A=\left[\sum_{j=i+1}^{m-1}(m-j)v_j-\sum_{j=-(m-1)}^{i-1}(m+j)v_j-
iv_i\right]^2.$$
Let us calculate the correlation between $u_0^r$ and $u_1^r$.
Only keeping the correlations between nearest neighbor, we have
\bqs
&&\left\langle\left[\sum_{j=1}^{m-1}(m-j)v_j-\sum_{j=-(m-1)}^{-1}(m+j)v_j\right]
^2\right\rangle\\
&=&2\left\langle\sum_{j=1}^{m-1}(m-j)^2v_j^2+2\sum_{j=1}^{m-2}
(m-j)(m-j-1)v_jv_{j+1}\right\rangle\\
&=&(m-1)^2\overline{v}_1^2+\overline{v}_{m-1}^2\\
&&+\sum_{j=1}^{m-2}[(m-j)\overline{v}_j-(m-j-1)\overline{v}_{j+1}]^2\\
&&+2a_2\epsilon\sum_{j=1}^{m-2}(m-j)(m-j-1)\overline{v}_j\overline{v}_{j+1}\\
&\equiv&A_1\overline{v}_0^2.
\eqs
So
\beq
\frac{\langle 2u_0^ru_1^r\rangle }{\langle {u_0^r}^2+{u_1^r}^2\rangle}=
\frac{A_1-m^2}{A_1+m^2}
\label{eq:thecor}
\eeq

Obviously, when $m=1$, the above ratio is $-1$, because $u_0^r=-u_1^r$
at all time.  For elastic particles,
the ratio can be calculated analytically to be $-\frac{1}{2m-1}$.  The
inelasticity changes this dependence.  Let us call $2m$ the `cluster size',
since it corresponds to the usual practice of defining a `cluster' then
separate the motion of particles into mean flow and fluctuations.

From expression (\ref{eq:thecor}), we see that when the cluster is large
enough, $A_1$ can be big
comparing to $m^2$, then the correlations
between velocity fluctuations can be big.

\subsection{Numerical results}

We carry out numerical simulations to investigate the statistical steady
state of the system.  Here we compare the numerical results with the
above theory describing the motion in the center of mass frame.

\subsubsection{Quasi-elastic situations}

First let us look at the quasi-elastic situations, i.e.\ very small $\epsilon$.
Before testing the profile of $v_i$'s, we exam the crucial assumption, 
(\ref{eq:a1}).

\begin{figure}
\narrowtext
\epsfxsize=\hsize
\vspace{-.2in}
\epsfbox{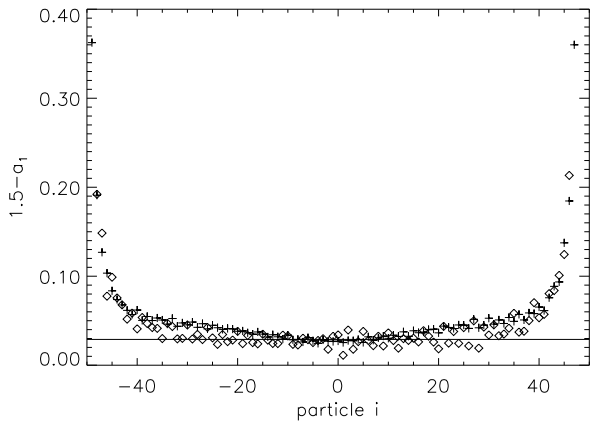}
\caption{Numerical results for $1.5-a_1$ for a high density system with
$100$ particles averaged over $2\times 10^9$ collisions.
$a_1$ is defined in Equation~(\ref{eq:a1}).
The $\diamond$ is for $r=0.995$; the  $+$ is for $r=0.95$.  The
line is for $1.5-a_1=0.029$ from the fit in Figure~\ref{fg:beps}.}
\label{fg:a1}
\end{figure}

Now let us look more sharply at the data.
To find $a_1$, we do a very accurate determination of the ratio of
averages from the left and the right hand sides of equation
(\ref{eq:a1}). This equation is then solved at each $i$-value
to find a local value of $a_1$. The result  is shown in Figure~\ref{fg:a1}.
The theory is right if  $a_1$ is independent of $i$ and wrong if it has an
important $i$-dependence. The figure seems to show that there is an
excellent fit for the smaller value of $\epsilon$, and a bad fit for the
larger.

From equation (\ref{eq:b}) we see that the important combination
determining the properties of the profile of $\bar{v}$ is $3-2a_1$.
But $a_1$ is very close to $1.5$, as shown in Figure~\ref{fg:a1}.
Then the $a_1$ effect  changes the prefactor in equation (\ref{eq:b})
from $3$ to $3-2a_1$, i.e.
by a factor of $50$.  The velocity correlations renormalize
$\epsilon$, and reduce the energy dissipation.

Also $a_1$ is essentially a local correlation effect originated from the
inelastic collisions.
For an elastic system with comparable inhomogeneity, there is also a
correction to the
factor $-\frac{1}{2}$ in Equation~(\ref{eq:a1}), but the correction
is usually
an order of magnitude smaller than the effects we are seeing here.

\begin{figure}
\narrowtext
\epsfxsize=\hsize
\vspace{-.2in}
\epsfbox{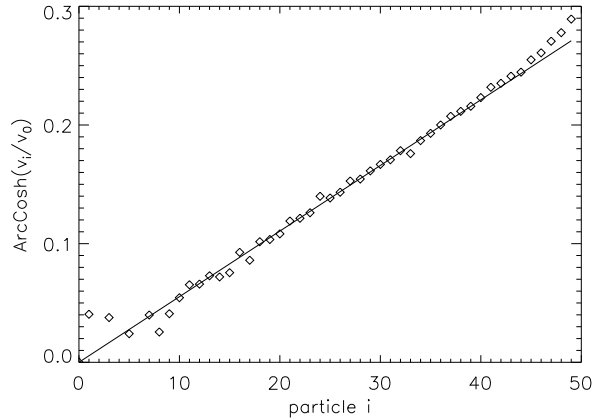}
\caption{Fit to a hyperbolic cosine curve of the profile of the
$\overline{v}_i$ for a
high density system of 100 particles and $r=0.9995$.
To check equation (\ref{eq:pro2}), we take the inverse of the hyperbolic
cosine
of $\bar{v}_i/\bar{v}_0 $ and plot the result as a function of $i$.
The straight line indicates a fit to the theory. In the theory, the slope
is proportional to the square root of $\epsilon$.
Here the slope is $0.0055$, which is equal to the square root
of $0.06\epsilon$.}
\label{fg:cosh}
\end{figure}

\begin{figure}
\narrowtext
\epsfxsize=\hsize
\vspace{-.2in}
\epsfbox{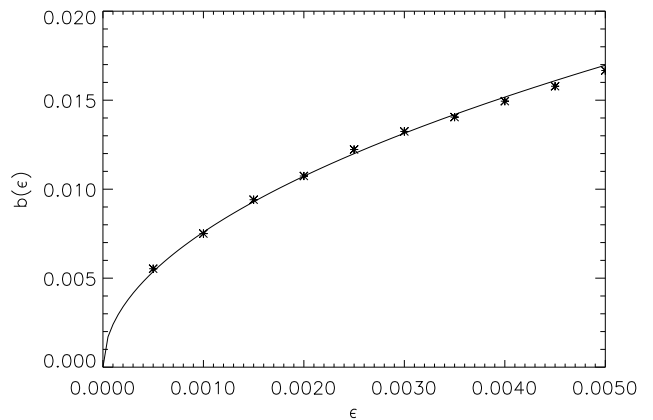}
\caption{The $\epsilon$ dependence of $b$ for systems with $N=100$.
The curve is the theoretical fit, the
square root of $0.058\epsilon$. (See Equation~(\ref{eq:pro2}).)}
\label{fg:beps}
\end{figure}

A test of (\ref{eq:pro2}) is showed in Figure~\ref{fg:cosh}.
Analysis like this permits the determination of the slope  like
the one in Figure~\ref{fg:cosh} as a function of $\epsilon$. We have called
this slope $b$.
Figure~\ref{fg:beps} shows that the numerical values give an $\epsilon$
dependence for $b$
which fully supports the theory.  However, notice that all this
analysis applies to very small values of $\epsilon$.
The next section considers more inelastic situations.

\subsubsection{Stronger inelasticity regime}

We look at smaller $r$'s.  To avoid inelastic collapses, we limit our
$r$ to be greater than $r_c$.  For a system of $100$ particles with
extremely high density,
$r_c\approx 0.95$.

When $r$ gets smaller, there is a cluster of particles moving around the
center of the pipe, all with about
the same velocity.  The system is in a state far away from equilibrium.
Also, it is very
nonuniform---the particles around the center are highly correlated while
those near the
boundaries move independently; the energy flux is strong near the end
walls, but rather
weak inside the system.  As a consequence, the PDF's of quantities
change significantly
from particles near the center to those near the boundaries,
e.g.\ the PDF's of $u_i$'s, though there is no big change in
the PDF's of $v_i$'s.

\begin{figure}
\narrowtext
\epsfxsize=\hsize
\epsfbox{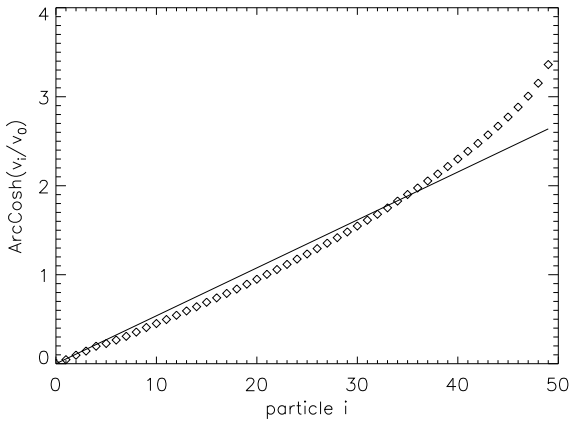}
\caption{Fit to a hyperbolic cosine curve of the profile of
$\overline{v}_i$ for a high density
system of 100 particles and $r=0.95$. This is a higher-$\epsilon$ analog of
Figure~\ref{fg:cosh}.  The straight line corresponds
to a hyperbolic
cosine profile-curve, and its slope is $0.054$, a value extrapolated from the
expression for quasi-elastic cases (Fig.~\ref{fg:beps}). However, the
straight-line
fit is not very good, especially near the boundary.}
\label{fg:dech}
\end{figure}

\begin{figure}
\narrowtext
\epsfxsize=\hsize
\epsfbox{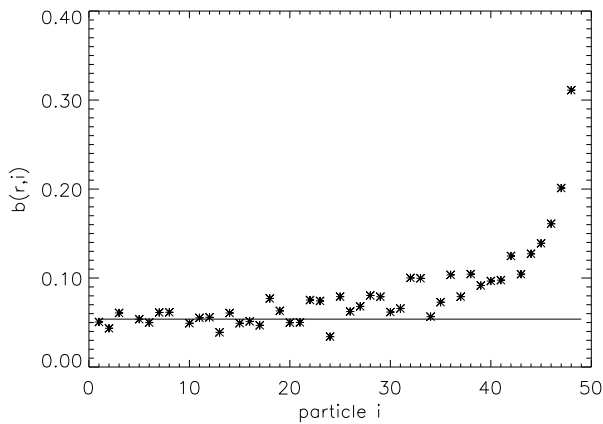}
\caption{The position dependence of the prefactor $b$ in a high density
system with $100$ particles
and $r=0.95$.  The line is $b=0.054$ extrapolated from the quasi-elastic
cases.}
\label{fg:br}
\end{figure}

Figure \ref{fg:dech} once again plots a quantity which should be linear in
$i$ if the theory,
equations (\ref{eq:pro2}) , is right.   Now, for this larger values of
$\epsilon$, there are  substantial variations in slope.  It appears that
the theory does not
apply for the fifteen particles nearest to each of the boundaries and that
it might have
small troubles elsewhere. This discrepancy is also shown when we plot the
slope,
calculated from doing numerical derivatives on Figure~\ref{fg:dech}
to give $b$ as a function of $i$ .  This plot is given as
Figure~\ref{fg:br}.

The discrepancy between the theory and numerical results for strong
inelasticity is not surprising.  Though taking into account the correlations
between fluctuations, the theory is still based on concepts of 
conventional fluids---no internal structures are considered.  However,
when inelasticity is strong, the dynamics is affected by intrinsic
structures of the collection of the particles, 
and the whole system may belong a different phase\cite{zhou}.  
A satisfactory theory must incorporate this feature.

\begin{figure}
\narrowtext
\epsfxsize=\hsize
\vspace{-.2in}
\epsfbox{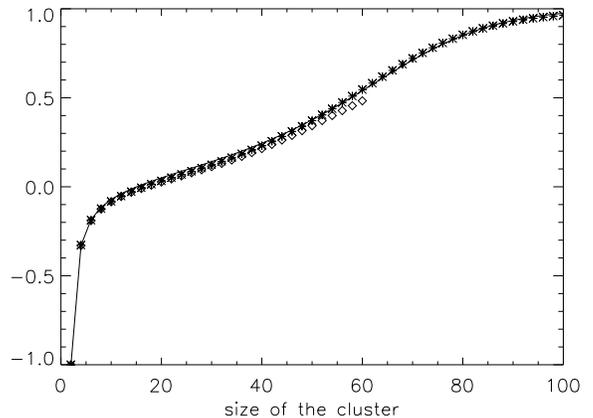}
\caption{The cluster size dependence of the ratio (\ref{eq:thecor}).  The system is
in a low density regime, with $r=0.94$.  $\ast$ is from
time average results of a simulation with $100$ particles and $\diamond$ is from
a simulation with $60$ particles, 
and the curve is from (\ref{eq:thecor}) with $a_2=0.6$.}
\label{fg:csize}
\end{figure}

Now let us look at the velocity correlations.  Only the nearest neighbor
correlation (Equation~(\ref{eq:a2})) is considered.  The theory leads to
the expression (\ref{eq:thecor}) of the correlation between $u_0^r$ and
$u_1^r$, which is independent of system sizes or boundary conditions.
To test this expression, we calculate numerically this correlation
with respect to different cluster sizes, i.e.\ different $m$,
with (\ref{eq:thecor}) and with the profile
of $\overline{v}_i$ calculated numerically.  The comparison between theory
and numerical result
is showed in Figure~\ref{fg:csize}.  We see the correlation increases with
increasing cluster size.
The comparison is the best for $a_2=0.6$.
When the cluster size is big enough, most part of the total motion belongs
to the correlated motion.  We want to point out that this curve is 
independent of boundary conditions.  Also for systems with different sizes,
we get sections of different length from this same curve, as shown in the
figure.

We want to point out that the major point of Figure~\ref{fg:csize} is to
demonstrate that part of the dynamics, the motion in the center of mass
frame, is independent of boundary conditions and system sizes.  The
agreement between theory and numerical results can not be viewed as a
strong support for the details of the theory because the profile of $\overline{v}_i$
is from numerical calculations, rather than (\ref{eq:pro2}), also the
value $a_2=0.6$ is a fitting parameter.  The theory captures some 
qualitative features of the dynamics, but is still incomplete.

\section{Motion of the center of mass}

\begin{figure}
\narrowtext
\epsfxsize=\hsize
\vspace{-.2in}
\epsfbox{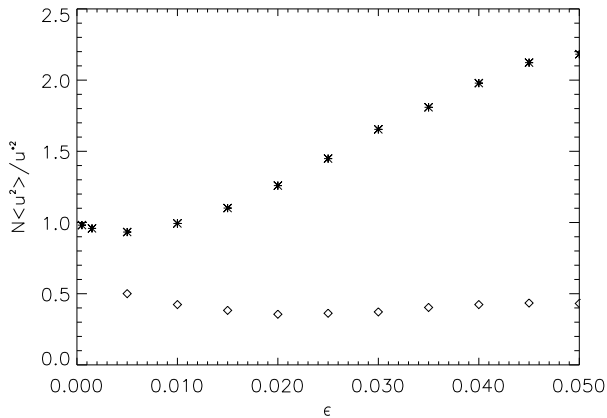}
\caption{Test of (\protect\ref{eq:u}) for two boundary conditions
for high density systems with $N=100$.
The ratios are all around $1$, as we expect from our order of magnitude
argument.
The $\ast$ is for the Boltzmann boundary condition, and the  $\diamond$
is for the fixed speed condition.  The motion of the center of mass
depends strongly on the boundary conditions.}
\label{fg:u}
\end{figure}

Because the total momentum of the system can be only changed by the collisions
between the outermost particles and the walls, and the motion of the outermost
particles
is close to that of a elastic system, the motion of the center of mass
should also be
close to that of a elastic system.  For a elastic system,
\beq
\langle u^2\rangle =\langle (\sum_i u_i/N)^2\rangle = {u^*}^2/N
\label{eq:u}
\eeq
where $u^*$ is the $rms$ speed of the outermost particle.  From
Figure~\ref{fg:u} we see this estimate is
about right, though the numerical factor must be calculated from detailed
distributions.  The result also seems sensitive to $\epsilon$.  This is because
the PDF for the velocity of the outermost particle is more skewed for higher
value of $\epsilon$, and so the ratio between $u^*$ and
the momentum transfered into
the system from the wall depends on $\epsilon$.

Notice that the motion of the center of mass depends strongly on the
boundary condition.

Suppose the motion of particles in the center of mass frame is independent of
the motion of the center of mass itself, i.e.\ $u$ is uncorrelated to $v_i$'s,
then
$$\langle u_i^2\rangle=\langle {u_i^r}^2\rangle+\langle u^2\rangle.$$
Simulations show that
the profile of $\overline{v}_i$ is nearly independent of boundary
conditions, so is the motion of the system in the center of mass frame.
However, $\langle u^2\rangle$
depends sensitively on the boundary conditions, and so does the motion of the
particles in the lab frame, i.e.\ the profile of $\langle u_i^2\rangle$
(Fig.~\ref{fg:boundary}).

Due to the motion of the center of mass, the correlations between $u_i$'s
are enhanced, comparing to those between $u_i^r$'s.
$$\frac{\langle 2u_iu_{i+1}\rangle}{\langle u_i^2+u_{i+1}^2\rangle}
=\frac{2\langle u_i^ru_{i+1}^r\rangle+2\langle u^2\rangle}
{\langle {u_i^r}^2+{u_{i+1}^r}^2\rangle+2\langle u^2\rangle},$$
or
\beq
\frac{v_i^2}{\langle u_i^2+u_{i+1}^2\rangle}
=\frac{v_i^2}
{\langle {u_i^r}^2+{u_{i+1}^r}^2\rangle+2\langle u^2\rangle}=R_i.
\label{eq:rat2}
\eeq
The ratio, $R_i$, between random motion and total motion was defined by us in
Equation~(\ref{eq:rat}).

\subsection*{Behavior of the ratio $R_0$}

When $n\epsilon$ is small, we can expand expression (\ref{eq:rat2}), using
Equations (\ref{eq:pro2}), (\ref{eq:thecor}) and (\ref{eq:u}).  Keeping terms
linear in $\epsilon$, we have,
\bqa
-\log(R_0)=\frac{\epsilon}{2n}\big\{ \left[ (n-\alpha)^2+(n-1)\right] (3-2a_1) 
\nonumber\\
+2(n-1)(n-2)a_2/3\big\},
\label{eq:lineps}
\eqa
where $0<\alpha <1$.  From Equation (\ref{eq:lineps}) we see that
when $n\epsilon$ is small, $-\log(R_0)$ is proportional to $n\epsilon$.

\begin{figure}
\narrowtext
\epsfxsize=\hsize
\vspace{-.2in}
\epsfbox{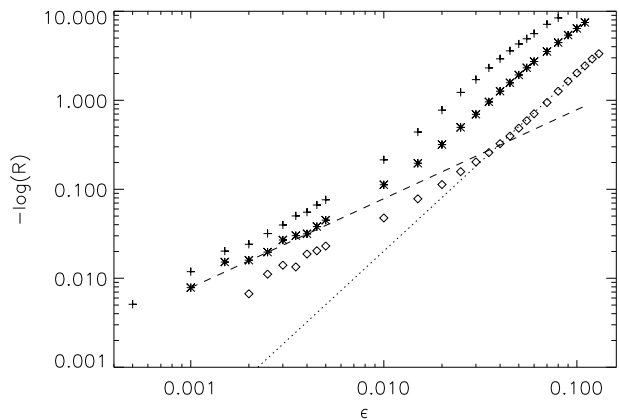}
\caption{The logarithm of the ratio (\protect\ref{eq:rat}) for $i=0$
versus $\epsilon$.  $+$ is for $N=100$, $\ast$ is for $N=70$, 
and  $\diamond$ is for $N=40$.  All three are for low density systems with
the Boltzmann boundary condition.  The dashed line indicates a
dependence $\log(R_0)\propto \epsilon$ and the dotted line indicates a
dependence $\log(R_0)\propto \epsilon^2$.}
\label{fg:mis}
\end{figure}

Numerical results of $-\log(R_0)$ are showed in
Figure~\ref{fg:mis}.  We do see that $-\log(R_0)$ is proportional to
$\epsilon$ for very small $\epsilon$.  However, when $\epsilon$ is big,
where we expect strong nonlinear effects, 
it is proportional to $\epsilon^2$.

\begin{figure}
\narrowtext
\epsfxsize=\hsize
\vspace{-.2in}
\epsfbox{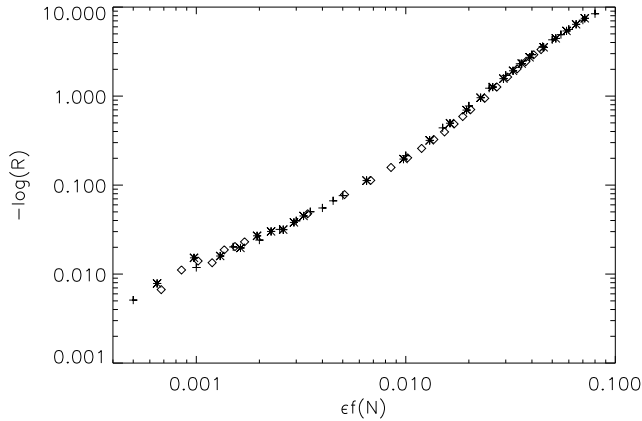}
\caption{The curves shown in Figure \ref{fg:mis} can be shifted to overlap
by changing the $x$-axis from $\epsilon$ to $\epsilon f(N)$, where $f(N)$ is
a function of the total number of particles in the system.  $f(100)=1$. 
Three curves are shown, they are all for low density systems with 
Boltzmann boundary conditions.  $+$
is for $N=100$, $\ast$ is for $N=70$ and $\diamond$ is for $N=40$.}
\label{fg:hit}
\end{figure}

\begin{figure}
\narrowtext
\epsfxsize=\hsize
\vspace{-.2in}
\epsfbox{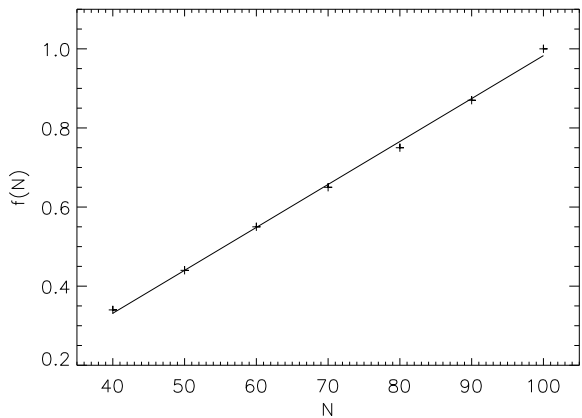}
\caption{The function $f(N)$ from Figure \ref{fg:hit}.  $f(N)$ is proportional
to $N$ with a small adjustment due to boundary effects.}
\label{fg:corre}
\end{figure}

As we argued in Section II, there are two important combinations of $N$
and $\epsilon$.  The product $N\sqrt{\epsilon}$ describes how temperature
decays towards the center of the system, which agrees excellently with the
numerical results when $\epsilon$ is very small.  However, 
Equation~(\ref{eq:lineps}) shows that in this limit, only the product
$N\epsilon$ appears in the final expression for $R_0$.  This seems to
suggest that $R_0$, i.e., the degree of the coherence of the 
particles' motion,
is determined by the product $N\epsilon$ (Fig.~\ref{fg:hit} and 
Fig.~\ref{fg:corre}).  These two figures exhibit rather interesting features
of the dynamics\cite{leo},
though we do not have a satisfactory understanding of them.

\section{Conclusion}

In this paper, we investigated the steady state of a forced granular system
in a thin pipe.
Correlations between velocities of granular particles are shown to be
important for
a correct understanding of such systems.  For systems in quasi-elastic
regime,  correlation
is small, but not negligible because the deviation from equilibrium is also
small.  For
systems with stronger inelasticity, correlation is crucial for a correct
theory.  Our theory describes the dynamics satisfactorily in the quasi-elastic
limit.  For stronger inelasticities, numerical results show quite interesting
behaviors of the system, however, our theoretical understanding is only
qualitative at this stage.

Characteristicly for granular systems,
fluctuations are important at all scales, enhanced by the combined
effects of momentum conservation and non-uniformity.  Also, the
separation between fluctuations and mean flow is quite nontrivial.
Because if the mean flow is an average of a collection of particles,
the correlations between the fluctuations of velocities can be big if the 
collection is big.

An important issue is the existence of a universal description, which is
not common for nonequilibrium systems.  The separation of the dynamics
into motion in the center of mass frame and the motion of the center of mass
itself is quite suggestive.  

The motion of the center of mass can not be universal.  Momentum conservation
decides that the velocity of the center of mass can be changed only by the
interaction between particles and external effects.  So it depends sensitively
on the details of boundary conditions, as shown in the paper, 
and can not be universal.

This is true for both elastic systems and inelastic ones.  However, in elastic
systems, every mode has the same strength due to equal-partition law of the
energy.  The motion of the center of mass is just one mode out of $Nd$ modes,
and its effect is negligible for a macroscopic system.  In a dissipative system,
on the other hand, being the only conserved mode, it can dominate over all
other modes.  Consequently, a universal description does not exist for the 
dynamics as a whole.

Still, if we look at the other $N-1$ modes which are perpendicular to this
non-universal mode, we may discover some universal features.  The
independence of the motion in the center of mass frame on the boundary
conditions and system sizes is a hint that this part of the dynamics may be
universal.  Further study is being carried out.

The thin pipe model used here simplifies greatly both the numerical and
analytical
calculations.  The low density version of it may not have higher
dimensional analogies,
where the sequence of particles is necessarily broken.  However, the high
density version
can be modified for a higher dimension situation, where the sequence can be
kept.

\section*{Acknowledgments}

This work is a continuation of the endeavor to understand the dynamics of
granular systems carried out in the Chicago group led by Prof.\ Leo Kadanoff.
Prof.\ Kadanoff formulated the pipe model as a nontrivial extension to the 
one-dimensional model.  I want to thank him for many helps and encouragements
during this work.
I am also grateful to E. Ben-Naim and D. Rothman for interesting discussions.  This
work was supported by the National Science Foundation under 
Awards \#DMR-9415604
and in part by the  MRSEC Program of the NSF under award number \#DMR-9400379.

\end{multicols}


\begin{thebibliography}{99}
\bibitem{leo}
Along with many others, this observation is due to Leo Kadanoff's insight.
\end{thebibliography}
\end{document}